# Going from classical to quantum description of bound charged particles
# Part 2: Implications for the atomic physics

A. L. KHOLMETSKII[1]([*]), T. YARMAN[2] and O. V. MISSEVITCH[3]

[1]*Department of Physics, Belarus State University, Minsk, Belarus*
[2]*Okan University, Istanbul, Turkey & Savronik, Eskisehir, Turkey*
[3]*Institute for Nuclear Problems, Minsk, Belarus*

**Summary**
This paper is the continuation of the analysis of bound quantum systems started in part 1 (T. Yarman, A.L. Kholmetskii and O.V. Missevitch. Going from classical to quantum description of bound charged particles. Part 1: basic concepts and assertions), which is based on a novel approach to the transition from classical to quantum description of electrically bound charges, involving the requirement of energy-momentum conservation for the bound electromagnetic (EM) field, when the EM radiation is forbidden. It has been shown that the modified expression for the energy levels of hydrogenic atoms within such a pure bound field theory (PBFT) provides the same gross and fine structure of energy levels, like in the standard theory. At the same time, at the scale of hyperfine interactions our approach, in general, does evoke some important corrections to the energy levels. Part of such corrections, like the spin-spin splitting in the hydrogen atom, is less than the present theoretical/experimental uncertainties in the evaluation of hyperfine contributions into the atomic levels. But the most interesting result is the appearance of a number of significant corrections (the 1*S*-2*S* interval and 1*S* spin-spin interval in positronium, 1S and 2S-2P Lamb shift in light hydrogenic atoms), which improve considerably the convergence between theoretical predictions and experimental results. In particular, the corrected 1*S*-2S interval and 1S spin-spin splitting in positronium practically eliminate the existing up to date discrepancy between theoretical and experimental data. The re-estimated classic 2*S*-2*P* Lamb shift as well as ground state Lamb shift in the hydrogen atom lead to the proton charge radius $r_p$=0.841(6) fm (from 2*S*-2*P* Lamb shift), and $r_p$=0.846(22) fm (from 1S Lamb shift), which perfectly agrees with the latest estimation of proton size via the measurement of 2*S*-2*P* Lamb shift in muonic hydrogen, i.e. $r_p$=0.84184(67) fm. Finally we consider the decay of bound muons in meso-atoms and achieve a quantitative agreement between experimental data and the results obtained through our approach.

## 1. Introduction

In part 1 [1] we analyzed the problem of transition from classical to quantum description of electrically bound particles, taking into account a substantial transformation of the structure of their EM field, when the particles are described either in the classical, or in the quantum way. The latter is related to the well-known fact that the wave-like bound particles in the stationary energy states do not emit EM radiation, whereas the classical charges undergoing an orbital motion must radiate. Hence, in order to provide a consistent transition from classical to quantum description of such systems, we explored the pure bound field classical electrodynamics (CED), where a motion of charges is described in a classical way, but their EM radiation is prohibited. Based on the energy-momentum conservation law for such charges, we have derived the Hamilton functions for the one- and two-body problems, which were considered as a prototype for the construction of corresponding Hamilton operators, being applied to the one-body and two-body problems in the atomic physics. Further on we have shown that the modified Dirac-Coulomb (DC) equation for quantum one-body problem and the modified Breit equation for the two-body problem yield the same gross as well as fine structure of energy levels for hydrogenic atoms, like in the conventional approach. In addition, we have demonstrated that the PBFT correction to spin-spin in-terval in the hydrogen and heavier atoms is much less than the present calculation uncertainty. At the same time, at the range of hyperfine contributions our approach, in general, evokes some important corrections to the energy levels, which are discussed in the present paper. We show that the corrections of PBFT to the fine structure (having the order of magnitude of hyperfine interactions) are significant only for 1$S$ states of hydrogenic atoms (sub-section 2.1). The correction brought by the PBFT to the hyperfine spin-spin interaction occurs significant for positronium, and allows eliminating the available disagreement between calculated and experimental data (sub-section 2.2). In sub-section 2.3 we analyze the PBFT corrections to the Lamb shift in light hydrogenlike atoms. In section 3 we show that the corrections brought by PBFT to the common results practically eliminate the available up to date discrepancy between theory and experiment for the 1$S$-2$S$ interval and spin-spin splitting in positronium, classic Lamb shift and ground state Lamb shift in hydrogen. In particular, we derive the proton charge radius $r_p$=0.841(6) fm (from 2$S$-2$P$ Lamb shift) and $r_p$=0.846(22) fm (from 1$S$ Lamb shift), which perfectly agrees with the latest experimental result [2]. In section 4 we discuss the physical meaning of the scaling transformation $\mathbf{r} = \mathbf{r'}/b_{mn}b_{Mn}\gamma_{mn}\gamma_{Mn}$, which has been applied to the solution of modified Breit equation without external field in [1]. In particular, we show that this scaling transformation can imply the change of proper time rate of bound particles as the function of their binding energy. Applying this result to re-calculation of the bound muon decay rate versus the atomic number $Z$, we reached much better correspondence of the corrected results to the available experimental data for muonic atoms in comparison with the standard predictions. Finally, section 5 contains a conclusion.

---
[*] E-mails: khol123@yahoo.com

## 2. - Hyperfine contributions to the energy levels for light hydrogenlike atoms

Last decades an appreciable progress has been achieved in physics of light hydrogenlike atoms, both in the theory and experiment. The measuring precision of the energy structure and uncertainty in its calculation now approach in some cases $10^{-13}$ [3-5]. Theoretical progress in QED calculations of energy levels in hydrogenlike atoms at the range of hyperfine interactions is hampered by the uncertainties of nuclear structure contribution, and the available precise experimental data allow to reverse the problem and to estimate nuclear characteristics by comparing theoretical and experimental results. By such a way, measuring the Lamb shift, one can extract the value of proton charge radius; the comparison of calculated and measured deuteron-hydrogen isotope shift yields the estimation of deuteron matter radius, etc.

There is a quite satisfactory agreement between theoretical predictions and experimental data regarding the light hydrogenic atoms, as well as in their comparison with the results in elementary particle physics. At the same time, there remain some deviations between theory and experiment, which cannot be attributed to any uncertain factors. In particular, in the determination of energy levels in light hydrogenic atoms, there are few clear discrepancies between theory and experiment, where the value of deviation ($\Delta$) substantially exceeds a corresponding uncertainty $\sigma$ (both theoretical and experimental):

- 1$S$-2$S$ interval in positronium ($\Delta/\sigma$=3.0);
- 1$S$ hyperfine interval in positronium ($\Delta/\sigma\approx$2.0) [5];
- proton charge radius $r_p$ derived from the classic 2$S$-2$P$ Lamb shift and the ground state Lamb shift in hydrogen systematically exceed the value of $r_p$ obtained from particle physics ($\Delta/\sigma$ varies from 2 to 5 according to different estimations [3, 6]);
- proton charge radius $r_p$ derived from the Lamb shift data on hydrogen differs from that of latest experiment with muonic hydrogen more than 4 % [2].

In what follows, we will consequently analyze the fine structure corrections (sub-section 2.1), corrections to spin-spin interaction (sub-section 2.2), and corrections to the Lamb shift in light hydrogenlike atoms (sub-section 2.3) within the approach of PBFT.

*2'1. Correction to the fine structure.* – In this sub-section we continue to analyze the modified Breit equation (55) derived in part 1 [1] and reproduced below for the convenience:

$$\left[ -\frac{\hbar^2 \nabla_{r'}^2 b_{Mn}}{2m} - \frac{\hbar^2 \nabla_{r'}^2 b_{mn}}{2M} - \frac{Ze^2}{r'} + \frac{1}{b_{mn}b_{Mn}\gamma_{mn}^2\gamma_{Mn}^2}\left( -\frac{p_b^4}{8m^3 b_{mn}^3 c^2} - \frac{p_b^2}{8M^3 b_{Mn}^3 c^2} + U_b(\boldsymbol{p}_{bm}, \boldsymbol{p}_{bM}, r') \right) \right] \psi(r') = W'\psi(r'), \quad (1)$$

where

$$\boldsymbol{r} = \boldsymbol{r}'/(b_{mn}b_{Mn}\gamma_{mn}\gamma_{Mn}), \quad (2)$$

$$W' = W/(b_{mn}b_{Mn}\gamma_{mn}^2\gamma_{Mn}^2), \quad (3)$$

the operator $U_b(\boldsymbol{p}_{bm}, \boldsymbol{p}_{bM}, r)$ is equal to

$$U_b(\boldsymbol{p}_{bm}, \boldsymbol{p}_{bM}, r) = -\frac{\pi Z e^2 \hbar^2}{2c^2}\left(\frac{1}{b_{mn}^2 m^2} + \frac{1}{b_{Mn}^2 M^2}\right)\delta(r) - \frac{Ze^2}{2b_{mn}b_{Mn}mMr}\left(\boldsymbol{p}_{bm}\cdot\boldsymbol{p}_{bM} + \frac{\boldsymbol{r}\cdot(\boldsymbol{r}\cdot\boldsymbol{p}_{bm})\boldsymbol{p}_{bM}}{r^2}\right) -$$

$$\frac{Ze^2\hbar\gamma_{mn}\gamma_{Mn}}{4b_{mn}^2 m^2 c^2 r^3}(\boldsymbol{r}\times\boldsymbol{p}_{bm})\cdot\boldsymbol{\sigma}_m + \frac{Ze^2\hbar\gamma_{mn}\gamma_{Mn}}{4b_{Mn}^2 M^2 c^2 r^3}(\boldsymbol{r}\times\boldsymbol{p}_{bM})\cdot\boldsymbol{\sigma}_M - \frac{Ze^2\hbar\gamma_{mn}\gamma_{Mn}}{2b_{mn}b_{Mn}mMc^2 r^3}\left((\boldsymbol{r}\times\boldsymbol{p}_{bm})\cdot\boldsymbol{\sigma}_M - (\boldsymbol{r}\times\boldsymbol{p}_{bM})\cdot\boldsymbol{\sigma}_m\right) +$$

$$\frac{Ze^2\hbar\gamma_{mn}\gamma_{Mn}}{4b_{mn}b_{Mn}mMc^2}\left[\frac{\boldsymbol{\sigma}_m\cdot\boldsymbol{\sigma}_M}{r^3} - 3\frac{(\boldsymbol{\sigma}_m\cdot\boldsymbol{r})(\boldsymbol{\sigma}_M\cdot\boldsymbol{r})}{r^3} - \frac{8\pi}{3}\boldsymbol{\sigma}_m\cdot\boldsymbol{\sigma}_M\delta(r)\right]. \quad (4)$$

and the PBFT factors $b_{mn}$, $b_{Mn}$, $\gamma_{mn}$, $\gamma_{Mn}$ are determined to the accuracy $(Z\alpha)^2$ by eqs. (58a-d) of ref. [1]. Hereinafter we follow to the designations of ref. [1].

We remind that eq. (1) differs from the common Breit equation without external field written in the Schrödinger-like form [6], by the replacements

$$m \to b_{mn}m,\ M \to b_{Mn}M,\ U \to \gamma_{mn}\gamma_{Mn}U,\ \boldsymbol{E} \to \gamma_{mn}\gamma_{Mn}\boldsymbol{E},\ \boldsymbol{B} \to \gamma_{mn}\gamma_{Mn}\boldsymbol{B} \quad (5\text{a-e})$$

(compare with eqs. (50a-c) of ref. [1]). As we have shown in [1], these replacements reflect the modification of the structure of bound EM field for the system of bound charged with the prohibited radiation, when the wave equation (12) for the operator of vector potential is replaced by the Poisson-like equation (14).

We further show that the corrections of PBFT to fine structure of light hydrogenlike atoms with $m \ll M$ have the order of magnitude of $(Z\alpha)^6(m/M)$ and higher, and due to their scaling as $n^{-6}$ (or $n^{-5}$), they should be taken into account practically only for the ground states. In the case of positronium, the fine structure corrections occur significant not only due to the equality $m=M$, but also due to the PBFT correction to the annihilation term.

First of all, we extract from the Hamiltonian of eq. (1) the terms of fine interactions with the order $(Z\alpha)^4$ properly modified in PBFT, omitting at this stage the contribution due to spin-spin interaction. Hence, via eqs. (1) and (4) we obtain the operator of fine interaction in the form

$$(V_b(r))_{fine} = (V_b(r))_{rel} + (V_b(r))_{contact} + (V_b(r))_{s-o}, \quad (6)$$

where

$$(V_b(r))_{rel} = \frac{1}{b_{mn}b_{Mn}\gamma_{mn}^2\gamma_{Mn}^2}\left(-\frac{p_b^4}{8m^3 b_{mn}^3 c^2} - \frac{p_b^4}{8M^3 b_{Mn}^3 c^2}\right) \quad (7)$$

is the relativistic term,

$$(V_b(r))_{contact} = -\frac{1}{b_{mn}b_{Mn}\gamma_{mn}^2\gamma_{Mn}^2}\frac{\pi Z e^2 \hbar^2}{2c^2}\left(\frac{1}{m^2 b_{mn}^2} + \frac{1}{M^2 b_{Mn}^2}\right)\delta(r) \quad (8)$$

is the term of contact interaction, and

$$(V_b(r))_{s-o} = -\frac{Ze^2\hbar}{b_{mn}b_{Mn}\gamma_{mn}\gamma_{Mn}c^2 r^3}\left[\frac{(\boldsymbol{r}\times\boldsymbol{p}_{bm})\cdot\boldsymbol{\sigma}_m}{2m^2 b_{mn}^2} + \frac{(\boldsymbol{r}\times\boldsymbol{p}_{bM})\cdot\boldsymbol{\sigma}_M}{2M^2 b_{Mn}^2} - \frac{((\boldsymbol{r}\times\boldsymbol{p}_{bm})\cdot\boldsymbol{\sigma}_M - (\boldsymbol{r}\times\boldsymbol{p}_{bM})\cdot\boldsymbol{\sigma}_m)}{2b_{mn}b_{Mn}mM}\right]$$

(9)

is the term of spin-orbit interaction.

We point out that the operators (7)-(9) are presented in *r*-coordinates, whereas in the Hamiltonian (1) they should be expressed through $r'$− coordinates. In the latter case, each of the terms (7)-(9) can be calculated with taking into account of eq. (2), as well as the equalities

$$p_b^4(r') = b_{mn}^4 b_{Mn}^4 \gamma_{mn}^4 \gamma_{Mn}^4 p_b^4(r), \quad \delta(r) = b_{mn}^3 b_{Mn}^3 \gamma_{mn}^3 \gamma_{Mn}^3 \delta(r'). \tag{10-11}$$

Introducing the replacements (10-11) into eq. (7-9), and expressing the terms (8), (9) via the fine structure constant, we obtain the operator of fine interactions in $r'$-coordinates as follows:

$$(V_b(r'))_{fine} = \gamma_{mn}^2 \gamma_{Mn}^2 \begin{pmatrix} -\dfrac{p_b^4(r')b_{Mn}^3}{8m^3 c^2} - \dfrac{p_b^4(r')b_{mn}^3}{8M^3 c^2} + \dfrac{\pi(Z\alpha)b_{Mn}^2}{2m^2 c^2}\delta(r') + \dfrac{\pi(Z\alpha)b_{mn}^2}{2M^2 c^2}\delta(r') + \\ \dfrac{(Z\alpha)b_{Mn}^2}{2m^2 r'^3 c^2}(r' \times p_{bm}(r')) \cdot \sigma_m + \dfrac{(Z\alpha)b_{mn}^2}{2M^2 r'^3 c^2}(r' \times p_{bM}(r')) \cdot \sigma_M - \\ \dfrac{(Z\alpha)b_{mn}b_{Mn}}{2mMc^2 r^3}((r \times p_{bm}) \cdot \sigma_M - (r \times p_{bM}) \cdot \sigma_m) \end{pmatrix}. \tag{12}$$

Then we can rewrite eq. (1) in the form

$$\left[ -\dfrac{\hbar^2 \nabla_{r'}^2 b_{Mn}}{2m} - \dfrac{\hbar^2 \nabla_{r'}^2 b_{mn}}{2M} - \dfrac{Ze^2}{r'} + (V_b(r'))_{fine} \right] \psi(r') = \dfrac{W}{b_{mn} b_{Mn} \gamma_{mn}^2 \gamma_{Mn}^2} \psi(r'). \tag{13}$$

Now our immediate goal is to determine the fine structure corrections of PBFT in eq. (13), which may emerge in the order $(Z\alpha)^6$. To solve this problem, we need to determine factors $b_{mn}$, $b_{Mn}$, $\gamma_{mn}$, $\gamma_{Mn}$ to the order $(Z\alpha)^4$.

For this purpose, first of all, we find the classical coefficients $b_m$, $b_M$, and $\gamma_m$, $\gamma_M$ to the accuracy $c^{-4}$ based on their definitions (51a-d) of ref. [1]. The modification of factors $b_m$, $b_M$ issues from the two circumstances:
- the factors $\gamma_m$ and $\gamma_M$ in eqs. (51c-d) of ref. [1] are no longer adopted to be equal to unity, but they have to be determined to the accuracy $(v/c)^2$;
- in the expression for interaction EM energy (48) of ref. [1], the term of magnetic interaction energy is no longer ignored. Taking into account that for the bound EM field $\boldsymbol{B} = (\boldsymbol{v} \times \boldsymbol{E})/c$, we obtain for the circular motion in the classical two-body problem

$$\boldsymbol{B}_m \cdot \boldsymbol{B}_M = B_m B_M = \dfrac{v_m v_M}{c^2} E_m E_M = \dfrac{mM}{(m+M)^2} \dfrac{v_R^2}{c^2} \boldsymbol{E}_m \cdot \boldsymbol{E}_M.$$

Substituting this equality into eq. (48) of [1] and using the quantum mechanical definition of $\overline{U}$, we obtain

$$b_{mn} = \left(1 + \dfrac{\gamma_{Mn} \overline{U}}{mc^2}\right) = 1 - \dfrac{(Z\alpha)^2}{n^2}\left(1 + \dfrac{(Z\alpha)^2}{2n^2}\dfrac{m^2}{(m+M)^2}\right)\left(1 + \dfrac{mM}{(m+M)^2}\dfrac{(Z\alpha)^2}{n^2}\right)\dfrac{M}{(m+M)} \approx$$

$$1 - \dfrac{(Z\alpha)^2}{n^2}\dfrac{M}{m+M} - \dfrac{(Z\alpha)^4}{2n^4}\dfrac{m^2 M}{(m+M)^3} - \dfrac{(Z\alpha)^4}{n^4}\dfrac{mM^2}{(m+M)^3} \tag{14a}$$

to the accuracy $(Z\alpha)^4$, where we have used eqs. (51e-f) of [1], taking into account that $\overline{v_R^2} = \dfrac{(Z\alpha)^2}{n^2} c^2$, and putting $\left[1 - \dfrac{(Z\alpha)^2}{n^2}\dfrac{m^2}{(m+M)^2}\right]^{-1/2} \approx 1 + \dfrac{(Z\alpha)^2}{2n^2}\dfrac{m^2}{(m+M)^2}$ with the sufficient accuracy of calculations.

Similarly we determine the factor

$$b_{Mn} = 1 - \dfrac{(Z\alpha)^2}{n^2}\dfrac{m}{m+M} - \dfrac{(Z\alpha)^4}{2n^4}\dfrac{M^2 m}{(m+M)^3} - \dfrac{(Z\alpha)^4}{n^4}\dfrac{Mm^2}{(m+M)^3}. \tag{14b}$$

We point out that expanding the classical coefficients $b_{mn}$ and $b_{Mn}$ (eqs. (51a-b) of [1]) to the accuracy $(v/c)^4$, we adopt that the involvement of non-Coulomb interactions does not affect their values. Indeed, at the semi-classical level, as shown in ref. [8], such non-Coulomb interactions are exactly counteracted by corresponding change of the Coulomb interaction energy due to the proper variation of the radius of electron's orbit. Hence the resultant action of Coulomb and non-Coulomb interactions is equivalent to the Coulomb interaction with the appropriately adjusted electron's orbit. This observation concurrently implies that the overall change of the energy of semi-classical system "electron plus nucleus", for example, due to spin-orbit interaction exhibits as a proper change of kinetic energy of the orbiting electron and nucleus by the energy of non-Coulomb interaction [8]. Therefore, the taking into account of fine interactions does modify the factors $\gamma_{mn}$ $\gamma_{Mn}$ in the orders $(Z\alpha)^4$ and higher. In order to find the related corrections, it is convenient to use the relationships between the Lorentz factors and momenta of particles

$$\gamma_{mn}^{2} = \frac{1}{1 - p_b^2/m^2c^2}, \quad \gamma_{Mn}^{2} = \frac{1}{1 - p_b^2/M^2c^2},$$

where $p_b$ is the value of non-relativistic momentum of electron (nucleus), and to the order $(Z\alpha)^2$ the averaged value of $p_b^2$ is defined by eq. (60) of ref. [1].

With the involvement of fine interactions, this expression is modified as

$$\overline{p^2} = 2m_R \left( \overline{(W_{0n} + V_{fine}) + \frac{Ze^2}{r}} \right). \text{ Taking into account eq. (60) of [1], we obtain}$$

$$\gamma_{mn}^{2} = \frac{1}{1 - \frac{(Z\alpha)^2}{n^2}\frac{M^2}{(m+M)^2} - \frac{2m_R(V_b(r'))_{fine}}{m^2c^2}}, \quad \gamma_{Mn}^{2} = \frac{1}{1 - \frac{(Z\alpha)^2}{n^2}\frac{m^2}{(m+M)^2} - \frac{2m_R(V_b(r'))_{fine}}{M^2c^2}}. \quad (14\text{c-d})$$

Having obtained the coefficients $b_{mn}$ and $b_{Mn}$ (eqs. (14a-b)) and $\gamma_{mn}$, $\gamma_{Mn}$ (eqs. (14c-d)) we now in the position to calculate the fine structure corrections of PBFT to the order $(Z\alpha)^6$ on the basis of eq. (13). In particular, we find that

$$\frac{\overline{p^2}b_{Mn}}{2m} + \frac{\overline{p^2}b_{mn}}{2M} = \frac{\overline{p^2}}{2m_R} - m_R c^2 \frac{(Z\alpha)^4}{n^4}\frac{Mm}{(m+M)^2} - (V_b(r'))_{fine}\frac{(Z\alpha)^2}{n^2}\frac{2Mm}{(m+M)^2}. \quad (15)$$

Further, we determine the product $\dfrac{W}{b_{mn}b_{Mn}\gamma_{mn}^{2}\gamma_{Mn}^{2}}$, which for the $nS$ state is equal to

$$\frac{W}{b_{mn}b_{Mn}\gamma_{mn}^{2}\gamma_{Mn}^{2}} = W - \frac{m_R c^2 (Z\alpha)^4}{n^4}\frac{Mm}{(m+M)^2} + \overline{V}_f \frac{(Z\alpha)^2}{n^2} - \frac{m_R c^2 (Z\alpha)^6 5mM}{4n^6(m+M)^2} - \frac{m_R c^2 (Z\alpha)^6 m^2 M^2}{2n^6(m+M)^4}.$$

(16)

Here we have used eqs. (14a-d), and in the terms of the order $(Z\alpha)^4$ we put $W = W_{0n} = -\dfrac{m_R c^2 (Z\alpha)^2}{2n^2}$, while in the terms of the lower order $(Z\alpha)^2$ we presented $W = W_{0n} + \overline{V}_{fine}$.

In order to find the PBFT corrections for the term $(V_b(r'))_{fine}$, we can write it in the form

$$(V_b(r'))_{fine} = (V(r'))_{fine} + \delta V_{fine},$$

where $(V(r'))_{fine}$ represents the common operator of fine structure, and $\delta V_{fine}$ contains the specific corrections of PBFT determined by the substitution of factors $b_{mn}$, $b_{Mn}$, $\gamma_{mn}$, $\gamma_{Mn}$ in eq. (12). Since the operator $(V(r'))_{fine}$ itself has the order $(Z\alpha)^4$, it is enough to use the PBFT factors (14a-d) written to the accuracy $(Z\alpha)^2$. In particular, for $nS$-state of hydrogenlike atom, the straightforward calculations yield:

$$\delta V_{fine} = \frac{M^2 + m^2}{(M+m)^2}\frac{(Z\alpha)^2}{n^2}\overline{V}_f + \frac{2mc^2(Z\alpha)^6}{n^5(M+m)^5}\left(mM^4 - m^2M^3 + m^3M^2\right) - \frac{9}{8}\frac{mc^2(Z\alpha)^6}{n^6}\frac{mM^2(M^2+m^2)}{(M+m)^5}. \quad (17)$$

Substituting eqs. (14-17) into eq. (13), we obtain after lengthy, but straightforward calculations for $nS$-states:

$$\left[\frac{p^2}{2m_R} - \frac{Ze^2}{r'} + (V)_f\right]\psi(r') =$$

$$\left\{W - \frac{2mc^2(Z\alpha)^6}{n^5(M+m)^5}\left(2mM^4 + m^2M^3 + 2m^3M^2\right) + \frac{mc^2(Z\alpha)^6}{n^6}\frac{mM^2}{4(m+M)^3} + \frac{m_R c^2(Z\alpha)^6 m^2M^2}{2n^6(m+M)^4}\right.$$
$$\left. + mc^2\frac{(Z\alpha)^6}{2n^6}\frac{(M^2+m^2)mM^2}{(m+M)^5} + \frac{9}{8}\frac{mc^2(Z\alpha)^6}{n^6}\frac{mM^2(M^2+m^2)}{(M+m)^5}\right\}\psi(r').$$

(18)

Designating the term in the bracket of *rhs* of eq. (18) as $W_{common}$, we rewrite this equation in the short form

$$\left[\frac{p^2}{2m_R} - \frac{Ze^2}{r'} + (V)_f\right]\psi(r') = W_{common}\psi(r'),$$

which represents the common Breit equation without spin-spin term written in the Schrödinger-like form (see, *e.g.* [6]), expressed in $r'$-coordinates. Therefore, the PBFT correction to the atomic $nS$ state energy levels is equal to

$$\delta W_{PBFT} = W - W_{common} = \frac{2mc^2(Z\alpha)^6}{n^5(M+m)^5}\left(2mM^4 + m^2M^3 + 2m^3M^2\right) -$$
$$\frac{mc^2(Z\alpha)^6 m^2M^3}{n^6(m+M)^5} - \frac{15}{8}\frac{mc^2(Z\alpha)^6}{n^6}\frac{mM^2(M^2+m^2)}{(M+m)^5}.$$

(19)

One can see that for atoms with $m \ll M$, the correction (19) has the order of magnitude $mc^2(Z\alpha)^6 m/M$ and scales as $n^{-6}$ or $n^{-5}$. Therefore, there is no practical need to calculate this correction for $n \geq 2$ and $l \neq 0$.

In fact, the correction (19) is significant for $1S$ state only, and for hydrogen it is equal to

$$\delta W_{fine}^H(1S) = \frac{9}{8}mc^2(Z\alpha)^6\frac{m}{M} = 10.8 \text{ kHz} \quad (20a)$$

Note that this correction is positive, and it reduces the value of $1S$ state (which is negative). For $2S$ state of hydrogen we obtain from eq. (19) $\delta W_{fine}^H(2S) = 0.6$ kHz. Hence the correction to $1S$-$2S$ interval in hydrogen is equal to

$$\delta W_{fine}^H(1S - 2S) = 10.2 \text{ kHz}. \quad (20b)$$

Below the correction (20b) will be involved into the re-estimation of $1S$ Lamb shift in hydrogen.

For $1S$ state of muonium, this correction is about 0.1 MHz, and is few times less than the theoretical uncertainty in calculation of $1S$-$2S$ transition ($\approx$0.30 MHz [3]), and much less than the experimental uncertainty in the measurement of this interval (9.8 MHz [9]).

For $1S$ state of positronium ($m = M$), the correction (19) becomes

$$\delta W_{fine}^{Ps}(1S) = \frac{21mc^2(Z\alpha)^6}{128} = 3.07 \text{ MHz},$$

and for the $1S$-$2S$ interval, as follows from eq. (19),

$$\delta W_{fine}^{Ps}(1S - 2S) = 2.95 \text{ MHz}, \quad (20c)$$

which will be involved below into the re-estimation of $1S$-$2S$ interval in positronium.

Besides, for positronium the Breit potential includes the additional annihilation part (*e.g.*, [6, 10]), which in PBFT acquires the form

$$(V_b(r))_{ann} = \frac{1}{b_n^2 \gamma_n^4}\left(\frac{\pi(Z\alpha)}{m^2 b_n^2 c^2}\left(\frac{7}{6}(3 + \boldsymbol{\sigma}_+ \cdot \boldsymbol{\sigma}_-) - 2\right)\delta(r)\right),$$

where $\boldsymbol{\sigma}_+$ ($\boldsymbol{\sigma}_-$) belongs to positron (electron), and we designated $b_{mn} = b_{Mn} = b_n$, $\gamma_{mn} = \gamma_{Mn} = \gamma_n$. In $r'$-coordinates, due to eq. (11), this operator acquires the form

$$(V_b(r'))_{ann} = b_n^2 \gamma_n^2 \left( \frac{\pi(Z\alpha)}{2m^2c^2}(3 + \boldsymbol{\sigma}_+ \cdot \boldsymbol{\sigma}_-)\delta(r') \right).$$

We average this term with the wave function for $l=0$ [6]

$$\psi(0) = \frac{1}{\sqrt{\pi}} \left( \frac{Z\alpha m_R}{n^2} \right)^{3/2},$$

and taking into account that for positronium $b_n = \left(1 - \frac{(Z\alpha)^2}{2n^2}\right)$, $\gamma_{mn} = \left[1 - \frac{(Z\alpha)^2}{4n^2}\right]^{-1/2}$ with sufficient accuracy of calculations $(Z\alpha)^2$ (see eqs. (14a-d) for $m=M$), we derive in the orthopositronium case:

$$(W_b)_{ann} = \left( \frac{mc^2(Z\alpha)^4}{3n^3} \left(1 - \frac{3(Z\alpha)^2}{4}\right) \right).$$

Hence the correction of PBFT to annihilation term reads:

$$\delta W_{ann} = \frac{mc^2(Z\alpha)^6}{4n^5}. \qquad (21)$$

The correction (21) decreases the value of 1$S$-2$S$ interval in positronium by 4.53 MHz, and thus, it should be added to the fine structure correction (20c). A comparison of the theory and experiment for 1$S$-2$S$ interval in positronium will be done below in sub-section 3.1.

2'2. *Corrections to hyperfine spin-spin splitting of energy levels in leptonic atoms.* – In section 5 of ref. [1] we obtained the following expression for calculation of spin-spin interval within PBFT:

$$(W_b)_{s-s} = \left(1 - \frac{(Z\alpha)^2}{n^2} \frac{2mM}{(M+m)^2}\right) W_{s-s}, \qquad (22)$$

where $W_{s-s}$ is the spin-spin splitting calculated in the common approach, so that the term

$$\delta(W_b)_{s-s} = -\frac{(Z\alpha)^2}{n^2} \frac{2mM}{(M+m)^2} W_{s-s} \qquad (23)$$

represents the correction of PBFT. We mentioned that for the hydrogen atoms $\delta(W_b)_{s-s}$ <100 Hz and can be ignored at the present accuracy of calculation of spin-spin interval. However, for the leptonic atoms (positronium, muonium), where the nuclear structure effects do not appear, the value (23) exceeds both theoretical and experimental uncertainty, and further analysis of eqs. (22-23) is required.

Now it is important to remind that the energy $W_{s-s}$ contains the ratios of magnetic moment to mass both for the electron and the nucleus, which are determined experimentally by means of the Zeeman effect. Since in PBFT the operator of interaction of magnetic dipole with an external magnetic field is, in general, modified, the appropriate corrections to the measured values "magnetic moment/mass" ratio should be clarified, too.

As known, the operator of interaction of two bound particles (electron and nucleus) with the external magnetic field reads [5]:

$$V_{mag} = g_m \frac{e\hbar}{2m}(\boldsymbol{s}_m \cdot \boldsymbol{B}) - g_M \frac{Ze\hbar}{2M}(\boldsymbol{s}_M \cdot \boldsymbol{B}), \qquad (24)$$

where $g_m$, $g_M$ are the g-factors for bound electron and nucleus, correspondingly. Being added to the Breit operator (1), along with the PBFT corrections (5a-e), this operator acquires the form:

$$(V_b)_{mag} = \frac{1}{b_{mn}b_{Mn}\gamma_{mn}^2\gamma_{Mn}^2} \left[ g_m \frac{e\hbar}{2mb_{mn}} \gamma_{mn}\gamma_{Mn}(\boldsymbol{s}_m \cdot \boldsymbol{B}) - g_M \frac{Ze\hbar}{2Mb_{Mn}} \gamma_{mn}\gamma_{Mn}(\boldsymbol{s}_M \cdot \boldsymbol{B}) \right].$$

Averaging this operator with the Schrödinger wave-function $\psi(r)$, due to the normalization requirement

$$\psi(r) = (b_{mn} b_{Mn} \gamma_{mn} \gamma_{Mn})^{3/2} \psi(r') \qquad (25)$$

implied by the transformation (2), we obtain

$$\left(\overline{V}_b\right)_{mag} \equiv (W_b)_{mag} = b_{mn} b_{Mn} \gamma_{mn}^2 \gamma_{Mn}^2 \left[ g_m \frac{e\hbar b_{Mn}}{2m}(s_m \cdot B) - g_M \frac{Ze\hbar b_{mn}}{2M}(s_M \cdot B) \right], \qquad (26)$$

where $(W_b)_{mag}$ gives the Zeeman splitting of energy levels in PBFT framework. Herein in the averaging of $\left(\overline{V}_b\right)_{mag}$ we put $B(r) = const$, which is always fulfilled in the atomic scale.

Inserting eqs. (14a-d) into eq. (26), we obtain for $nS$-state

$$(W_b)_{mag} \approx \left(1 - \frac{(Z\alpha)^2}{n^2} \frac{2Mm}{(m+M)^2}\right)\left[W_{mag} - \frac{(Z\alpha)^2}{n^2} \frac{e\hbar}{2m}((g_m s_m - Z g_M s_M) \cdot B)\frac{m}{M+m}\right], \qquad (27)$$

(with sufficient accuracy $(Z\alpha)^2$), where $W_{mag}$ stands for the Zeeman splitting of energy levels, obtained via the averaging of common operator (24). For the sublevel $F=1$ used for the measurement of Zeeman effect, $s_m - s_M = 0$, and eq. (27) reads:

$$(W_b)_{mag} \approx \left(1 - \frac{(Z\alpha)^2}{n^2} \frac{2mM_Z}{(m+M_Z)^2}\right)\left(W_{mag} - \frac{(Z\alpha)^2}{n^2} \frac{(g_m - Z g_M)e\hbar}{2m}(s_m \cdot B)\frac{1}{M+m}\right), \qquad (28)$$

where we supply the mass $M$ by the subscript "Z" ("Zeeman effect"), in order to distinguish it from the mass $M$ in eqs. (22-23), which designates the mass of the nucleus in the measurement of spin-spin splitting. Thus, the magnetic moment to mass ratio of derived from the Zeeman splitting should be corrected in PBFT for each bound particle with taking into account of the relationship (28) between $(W_b)_{mag}$ and $W_{mag}$.

For the $1S$ state of hydrogen and heavier atoms, the analysis of this correction in the estimation of spin-spin interval is not practically important, because we have found in ref. [1] that term $(Z\alpha)^2 \frac{2mM}{(M+m)^2} W_{s-s}$ of eq. (23) itself is less than 100 Hz (at $Z=1$) and is many times smaller than the nuclear-structure contribution to the $1S$ hyperfine splitting

Considering the spin-spin splitting of $1S$ state of muonium, we can also ignore the correction to the ratio "magnetic moment/mass" for the electron, because one can show that it induces the PBFT correction of the order 100 Hz, which is much smaller than the present theoretical uncertainly in calculation of spin-spin interval in muonium (about 500 Hz [5]). Further, for muonium we can put with a high accuracy $g_m = g_M$ in the term containing $(Z\alpha)^2$. We also use the known fact that the magnetic moment to mass ratio for bound muon is determined with the best accuracy via the Zeeman effect in muonium [5], so $M_Z = M$, and $Z=1$ With these equalities eq. (28) yields

$$(W_b)_{mag}^{Mu} = \left(1 - \frac{(Z\alpha)^2}{n^2} \frac{2mM}{(m+M)^2}\right) W_{mag}^{Mu},$$

which shows that the PBFT correction to spin-spin interval (23) has exactly the same structure as PBFT correction to the "magnetic moment/mass" ratio for bound muon. As a result, both corrections exactly compensate each other, and we get

$$(W_b)_{s-s}^{Mu} = \frac{\left(1 - \frac{(Z\alpha)^2}{n^2} \frac{2mM}{(M+m)^2}\right)}{\left(1 - \frac{(Z\alpha)^2}{n^2} \frac{2mM_Z}{(m+M_Z)^2}\right)} W_{s-s}^{Mu} = W_{s-s}^{Mu}. \qquad (29)$$

For positronium we have quite different situation ($M=m$, but $M_Z \approx 10^3 m$, if the magnetic moment to mass ratio for electron is taken from the Zeeman splitting in hydrogen or heavier atoms), and the correction to spin-spin interaction (23) dominates over the correction to the "magnetic moment/mass" ratio for bound electron. Hence we use the correction of eq. (23) solely, which yields

$$(W_b)^{Ps}_{s-s} = \left(1 - \frac{(Z\alpha)^2}{2n^2}\right) W^{Ps}_{s-s}. \tag{30}$$

This equation will be applied in sub-section 3.2 for PBFT correction of spin-spin interval for 1S state of positronium and for comparison with modern experimental data.

*2'3. Corrections to the Lamb shift.* - The corrections of PBFT obtained above in sub-sections 2.1 and 2.2 originate from the appropriate modification of the Breit equation suggested in ref. [1]. Analyzing now radiative corrections to the atomic energy levels, we consider PBFT as a complementary to QED, so that any modifications in the core structure of QED are not implied. Thus only the modifications of PBFT in the input of QED expressions (5a-e) should be accounted for, along with the relationship (25) for non-relativistic wave functions. On the basis of these results we derive below the corrections to the Lamb shift $L$ for light hydrogenlike atoms, which emerge in PBFT. With sufficient accuracy of calculations, we further adopt the limit $M \to \infty$ (one-body problem).

It is known that the dominant terms of the Lamb shift arise due to a finite radius of the electron $\langle r^2 \rangle$, which continuously emits and absorbs virtual photons, as well as due to vacuum polarization.

The finite radius of the electron induces a deviation from the Coulomb potential [3]

$$\delta V_{\text{finite radius}} = \frac{1}{6}\langle r^2 \rangle \Delta U \approx \frac{\alpha}{3\pi}\ln(Z\alpha)^{-2}\frac{\Delta U}{m^2} = \frac{4}{3}\ln(Z\alpha)^{-2}\frac{Z\alpha^2}{m^2}\delta(r), \tag{31}$$

where $\Delta$ is the Laplacian. According to eqs. (5a), (5c) for the bound electron the mass $m$ is replaced by $b_{mn}m$, while the Coulomb potential $U$ is replaced by $U' = \gamma_{mn}\gamma_{Mn}U \approx \gamma_{mn}U$, where we can put $\gamma_{Mn}=1$ with sufficient accuracy of calculations; further we also put $b_{Mn}=1$ in the correction to the Lamb shift. Thus the latter equation acquires the form

$$\delta V'_{\text{finite radius}} = \frac{4\gamma_n}{3b_n^2}\ln(Z\alpha)^{-2}\frac{Z\alpha^2}{m^2}\delta(r) = \frac{\gamma_{mn}}{b_{mn}^2}\delta V_{\text{finite radius}}, \tag{32a}$$

where $\delta V_{\text{finite radius}}$ is defined by eq. (31).

The contribution due to vacuum polarization [3] is also proportional to $\Delta U/m^2$, and thus the latter equation remains in force for this correction:

$$\delta V'_{\text{polarization}} = \frac{\gamma_{mn}}{b_{mn}^2}\delta V_{\text{polarization}}. \tag{32b}$$

The total contribution $\delta V_{\text{total}}$ is defined as the sum $\delta V_{\text{total}} = \delta V_{\text{finite radius}} + \delta V_{\text{polarization}}$, so that for the total perturbation we get

$$\delta V'_{\text{total}} = \frac{\gamma_{mn}}{b_{mn}^2}\delta V_{\text{total}}. \tag{32c}$$

The correction to the energy level is given by the matrix element of the total perturbation (32c), where we need to take into account eq. (25), putting $b_{Mn}, \gamma_{Mn}=1$. Hence

$$\Delta W_{PBFT} = \langle nS|\delta V'_{total}|nS\rangle = b_{mn}^3 \gamma_{mn}^3 \left(\frac{\gamma_{mn}}{b_{mn}^2}\Delta E\right) = b_{mn}\gamma_{mn}^4 \Delta W = \gamma_{mn}^2 \Delta W,$$

where $\Delta E$ denotes the value of energy shift, obtained within QED, and we have used the equality $b_{mn}\gamma_{mn}^2 = 1$, followed from eq. (14a) and (14c) at $M \to \infty$.

Thus the Lamb shift at the given energy level corrected within PBFT, reads:
$$(L_{nlj})_{PBFT} = \gamma_{mn}^2 L_{nlj}, \quad (33)$$
where $L_{nlj}$ stands for the Lamb shift calculated in QED.

For the 2S-2P Lamb shift both levels have a principal quantum number $n=2$, and we get
$$(L_b)_{2S-2P} = \gamma_{m2}^2 L_{2S-2P} = L_{2S-2P}\left(1-(Z\alpha)^2/4\right)^{-1}. \quad (33a)$$
Thus the correction induced by PBFR to the 2S-2P Lamb shift is equal to
$$\delta L_{2S-2P} = (L_b)_{2S-2P} - L_{2S-2P} = L_{2S-2P}\left[\left(1-(Z\alpha)^2/4\right)^{-1}-1\right]. \quad (33b)$$
Numerically this value is equal to 13.8 kHz for hydrogen, which exceeds substantially the measured precision for Doppler-free two-photon laser spectroscopy [4].

Below we compare the experimental and theoretical values for the corrected 2S-2P Lamb shift in hydrogen (sub-section 3.3) and 2S-2P Lamb shift in He$^+$ (sub-section 3.4).

Eq. (33) is, in general, also applicable to the 1S Lamb shift $L_{1S}$ in hydrogenlike atoms. However, its direct measurement is impractical until effects of nuclear structure are known accurately enough. In order to eliminate the influence of these effects, the data at least of two measurements are involved: for hyperfine intervals in the ground state and metastable states (for example, for the 1S and 2S states). Since the bulk contribution to the Lamb shift scales like $n^{-3}$, then the difference $8(W_{hpf})_{2S} - (W_{hpf})_{1S}$ allows us canceling substantially various contributions caused by the short distance effects. However, the factors $\gamma_{mn}$ differ from each other for 2S and 1S states, and calculation of the corrected 1S Lamb shift $(L_b)_{1S}$ is not straightforward.

In order to introduce the PBFT corrections to the 1S Lamb shift, to be convenient for practical applications, one need to look closer at the typical methods for its theoretical estimation. In principle, the 1S Lamb shift could be extracted from the experimental data on the transition frequencies between the energy levels with different numbers $n$. One should emphasize that the intervals of gross structure are mainly determined by the Rydberg constant $R$. In order to disentangle measurement of the 1S Lamb shift from the measurement of the Rydberg constant, one can use the experimental data on two different intervals 1S-2S and $2S_{1/2}$-$8D_{5/2}$ of hydrogen [11]. Theoretically these intervals can be presented as [3]
$$E_{1S-2S} = \left(W_{2S_{1/2}}^{DR} - W_{1S_{1/2}}^{DR}\right) + L_{2S_{1/2}} - L_{1S_{1/2}}, \quad (34a)$$
$$E_{2S-8D} = \left(W_{8D_{1/2}}^{DR} - W_{1S_{1/2}}^{DR}\right) + L_{8D_{5/2}} - L_{2S_{1/2}}, \quad (34b)$$
where $W_{nl_j}^{DR}$ is the leading Dirac and recoil contribution to the position of the respective energy level (eq. (63) of ref. [1]).

The differences of the leading Dirac and recoil contribution in the *rhs* of eqs. (34a-b) are proportional to the Rydberg constant $R$ plus corrections of order $\alpha^2 R$ and higher. One can construct a linear combination of these intervals which is proportional to $\alpha^2 R$ plus higher order terms
$$E_{1S-2S} - \frac{16}{5}E_{2S-8D} = \left(W_{2S}^{DR} - W_{1S}^{DR}\right) - \frac{16}{5}\left(W_{8D}^{DR} - W_{2S}^{DR}\right) - L_{1S} + \frac{21}{5}L_{2S} - \frac{16}{5}L_{8D}. \quad (35)$$
Then the difference of the leading Dirac recoil contribution in the *rhs* of eq. (35) can be calculated with a high accuracy, due to the suppression factor $\alpha^2$, and it practically does not depend on the exact value of $R$. Hence the linear combination of the Lamb shifts in the *rhs* of eq. (35) does not depend on $R$, too. The bulk contribution to the Lamb shift scales as $1/n^3$ which allows using the theoretical value $L_{8D_{5/2}}$=71.51 kHz [3] without loss of accuracy. The 2S Lamb shift can be extracted from the data on the classic 2S-2P Lamb shift, so that
$$L_{1S} = \left[\left(W_{2S}^{DR} - W_{1S}^{DR}\right) - \frac{16}{5}\left(W_{8D}^{DR} - W_{2S}^{DR}\right) - \frac{16}{5}L_{8D}\right] - \left[E_{1S-2S} - \frac{16}{5}E_{2S-8D}\right] + \frac{21}{5}L_{2S}. \quad (36)$$

Herein in *rhs*, the first term in square brackets is computed theoretically, the second term in square brackets is determined experimentally, while the last term is extracted from the data on the classic 2*S*-2*P* Lamb shift. Within PBFT, the first computed term in the *rhs* of eq. (36) should be corrected by adding the fine structure correction (20b). Besides, one has to correct within PBFT the 2*S* Lamb shift, using the data on the classic 2*S*-2*P* Lamb shift. Hence the expression for the 1*S* Lamb shift for the hydrogen within PBFT acquires the form

$$(L_b)_{1S}[kHz] = L_{1S}[kHz] + 10.2 + \frac{21}{5}\delta L_{2S}[kHz]; \qquad (37)$$

the PBFT correction to 2*S* Lamb shift $\delta L_{2S}$ will be found below in sub-section 3.5, where the value (37) will be calculated.

**3. - Corrections of PBFT to the atomic energy levels and comparison with experiment**

In this section we analyze the hyperfine contributions into the energy levels of light hydrogenic atoms, where the discrepancy between theory and experiment exceeds the uncertainties in their determination, and apply the appropriate corrections of PFBT derived above. We show that the corrections of PBFT provide a perfect conformity between theoretical and experimental values for all parameters listed in the introduction section: 1*S*-2*S* interval in positronium (sub-section 3.1); spin-spin splitting in positronium (sub-section 3.2), proton charge radius derived from the classic 2*S*-2*P* Lamb shift (sub-section 3.3), proton charge radius derived from the ground state Lamb shift in hydrogen (sub-section 3.5). We also pay a separate attention to the 2*S*-2*P* Lamb shift in He$^+$ (sub-section 3.4).

*3'1. 1S-2S interval in positronium.* - Modern theoretical value of this interval is [5]
$$E^{Ps}_{1S-2S} = 1\ 233\ 607\ 222.2(6)\ \text{MHz}, \qquad (38)$$
and the most precise result of experimental measurements is as follows:
$$1\ 233\ 607\ 216(2)\ \text{MHz [12]}. \qquad (39)$$

One can see that the deviation between the values (38) and (39) more than three times larger than the uncertainty in measurement of 1*S*-2*S* interval.

Now we introduce the PBFT correction to 1*S*-2*S* transition as the sum of eqs. (20c) and (21):
$$(\delta W_b)^{Ps}_{total}(1S-2S) = \delta W_b^{Ps}(1S-2S) + (\delta W_b)_{ann} = 7.48\ \text{MHz}.$$
Hence the 1*S*-2*S* interval in positronium corrected in PBFT becomes
$$(E_b)^{Ps}_{1S-2S} = E_{1S-2S} - (\delta W_b)^{Ps}_{1S-2S} = 1\ 233\ 607\ 214.7(6)\ \text{MHz}, \qquad (40)$$
which already agrees with the experimental value (39).

*3'2. 1S spin-spin interval in positronium.* - In ref. [1] and in sub-section 2.2 of the present paper we have shown that the correction of PBFT to hyperfine spin-spin interaction occurs quite negligible for hydrogen and muonium, eq. (29). For positronium we derived eq. (30), which now will be used for the comparison with experimental data.

The theoretical value of hyperfine splitting of 1*S* state of positronium is [5]
$$W^{Ps}_{s-s} = 203\ 391.7(8)\ \text{MHz}, \qquad (41)$$
which does exceed the corresponding experimental data 203 389(2) [13] and 203 387(2) [14].

Eq. (30) allows us to compute the corrected PBFT value of hyperfine spin-spin interval in positronium, using the numerical value (41):
$$(W_b)^{Ps}_{s-s} = 203\ 386(1)\ \text{MHz}. \qquad (42)$$
This result is already in a good agreement with the experimental data.

*3'3. 2S$_{1/2}$-2P$_{1/2}$ Lamb shift in hydrogen.* - It is well known that the dominant problem of exact theoretical evaluation of the classic Lamb shift $L_{2S-2P}$ is the uncertainty arising from the proton charge radius $r_p$. Due to this reason many authors reverse the problem, and estimate $r_p$ from the obtained data on $L_{2S-2P}$ shift (see, *e.g.*, [5]). It is also known that the estimated value of $r_p$ via the measurement of classic Lamb shift systematically exceeds the magnitudes of $r_p$, obtained in the electron-proton scattering data and other methods for evaluation of $r_p$ in physics of elementary particles [7]. This prompted scientists to assume [15] that the uncertainties in estimation of $r_p$ in the experimental particle physics are significantly underestimated. However, the very recent estimation of proton charge radius via the measurement of 2S-2P Lamb shift in muonic hydrogen gives the value $r_p$=0.84184(67) fm [2], which is substantially lower than the CODATA value $r_p$=0.8768(69) fm [16]. It is also important that for muonic hydrogen the nuclear size effect contributes significantly (about 2 %) to the 2S-2P Lamb shift and thus this new value of $r_p$ can pretend to be the most precise result amongst all published.

Below we will show that the PBFT correction (33b) to the 2S-2P Lamb shift removes the exiting remarkable disagreement between the estimation of $r_p$ from the classic Lamb shift data and estimation for muonic hydrogen.

First we determine the factor $\gamma_{m2}$, which for hydrogen and muonic hydrogen atoms has the value $\gamma_{m2} = (1-\alpha^2/4)^{-1/2}$ =1.0000066. For muonic hydrogen, where the nuclear size effect contributes significantly to the total 2S-2P energy interval, the corrected Lamb shift (33b) with the factor $\gamma_{m2}$ computed right above does not practically affect the proton charge radius estimated in ref. [2]. In particular, using the parameterization (1) of ref. [2] for the 2S-2P energy difference, one can show that the correction (33b) influences the estimated proton size in the order of magnitude 10$^{-4}$ fm, which is below of the measurement uncertainty [2]. In contrast, for the hydrogen atom the correction (33b) and finite nuclear size effect have comparable values, and the proton charge radius derived with and without correction (33b) acquires a difference to be substantially larger than the measured/calculated uncertainty.

In order to estimate the proton charge radius from the classic Lamb shift, we use the parameterization

$$L_{2S-2P}(r_p) = A + Br_p^2, \qquad (43)$$

which is based on the known fact [3] that the term proportional to $r_p^2$ is additive. Here *A* and *B* are the coefficients, whose numerical values can be found via common calculation of 2S-2P Lamb shift in hydrogen [3] for different values of proton charge radius [17, 18]:

*A*=1057695.05 kHz, *B*=195.750 kHz/fm. (44a-b)

In the framework of PBFT, the eq. (43) is appropriately modified:

$$(L_{2S-2P})_{PBFT} = \gamma_{m2}^2 A + \gamma_{m2}^2 B(r_{PBFT})_p^2, \qquad (45)$$

where $(r_{PBFT})_p$ is the proton size predicted by PBFT. Equating (43) and (45), we derive the expression for the corrected proton size in the framework of PBFT:

$$(r_{PBFT})_p = \sqrt{\frac{A(1-\gamma_{m2}^2)}{B\gamma_{m2}^2} + \frac{r_p^2}{\gamma_{m2}^2}} \approx \sqrt{r_p^2 - (Z\alpha)^2 \frac{A}{4B}} \qquad (46)$$

with the sufficient accuracy of calculations. Putting in eq. (46)

$r_p$=0.876(6) fm (47)

(CODATA value [16]), and using the numerical values (44a-b), we obtain

$(r_{PBFT})_p = 0.834(6)$ fm.

This estimation is much closer to the proton size derived in ref. [2], than the CODATA value (47). At the same time, we recall that the value (47) incorporates the experimental data in both particle physics and atomic physics, and, in general, is less than the proton size derived from the classic Lamb shift solely. In particular, the modern data on 2S-2P Lamb shift in hydro-gen obtained by various authors within the common approach (see refs. [3, 5] and references therein) define the range of variation of the values of $r_p$ between 0.875 fm and 0.891 fm. Thus taking the midpoint $r_p$ =0.883 fm, we obtain

$$(r_{PBFT})_p = 0.841(6) \text{ fm}, \tag{48}$$

which exactly coincides with the new proton size [2] (i.e. $r_p$=0.84184(67) fm).

*3'4. $2S_{1/2}$-$2P_{1/2}$ Lamb shift in He$^+$.* - Modern computed value of this shift is equal to [3]

$$L^{He}_{2S-2P} = 14\,041.46(3) \text{ MHz}, \tag{49}$$

which after the correction (33b) becomes

$$\left(L^{He}_b\right)_{2S-2P} = \left(\gamma^{He}_{m2}\right)^2 L^{He}_{2S-2P} = 14\,042.21(3) \text{ MHz} \tag{50}$$

(where $\gamma^{He}_{m2}$=1.0000266 for He$^+$). The result of measurement of the Lamb shift by an anisotropy quenching method reported in [19], is

$$\left(L^{He}_{\exp}\right)_{2S-2P} = 14\,042.52(16) \text{ MHz}, \tag{51}$$

which disagrees with both estimations (49) and (50).

The discrepancy between the experimental value (51) and QED prediction (49) stimulated further experimental research of the 2S-2P Lamb shift in He$^+$. In course of their work the authors of ref. [19] redesigned a photon detector system to eliminate a residual polarization sensitivity of the photon detectors, which, in authors' opinion, distorted the result of the previous measurement (51). Having implemented this improvement, they reported in [20] a new result

$$\left(L^{He}_{\exp}\right)_{2S-2P} = 14\,041.13(17) \text{ MHz}, \tag{52}$$

which again is in disagreement with the alternative predictions (49) and (50).

Thus, the performance of new high precision experiments on the subject appears to be highly required.

*3'5. 1S Lamb shift in hydrogen.* - Having corrected the classic 2S-2P Lamb shift in PBFT, we are now in the position to complete the PBFT corrections to the ground state Lamb shift (37). To the accuracy sufficient for further calculations, we adopt that the term $\delta L_{2S}$ in eq. (37) is completely determined by the corresponding correction to the coefficient $A$ in eq. (44a) for 2S-2P Lamb shift. Thus, we put $\delta L_{2S} \approx \delta A = (\gamma_{mn}^2 - 1)A$. Hence $\delta L_{2S}$=58.9 kHz. Inserting this value into eq. (37), we obtain

$$(L_b)_{1S} = L_{1S} + 69.1 \text{ kHz}. \tag{53}$$

Our next goal is to estimate the proton charge radius derived from the ground state Lamb shift corrected by eq. (53) via the comparison of calculated and experimental data on 1S Lamb shift collected in Table 12.3 of ref. [3]. One should point out that major part of experiments for the measurement of the 1S Lamb shift in hydrogen has been carried out with standard radiofrequency method, whose data are rather widely scattered between the values 8 172 798 kHz and 8 172 874 kHz, with the typical measurement error about 30-50 kHz. In these conditions we select the result of the mentioned Table [3]

$$L_{1S} = 8\,172\,837(22) \text{ kHz}. \tag{54}$$

obtained with Doppler-free two-photon laser spectroscopy [4], which provides more accurate determination of both 2S-2P and 1S Lamb shifts in comparison with radiofrequency method.

The same work [3] presents the corresponding theoretical values $(L_{1S})$ for two different values of proton charge radius:

$$L_{1S} = 8\,172\,663(6) \text{ kHz } (r_p\text{=0.805(11) fm}) \text{ [17], and} \tag{55a}$$

$$L_{1S} = 8\,172\,811(14) \text{ kHz (for } r_p\text{=0.862(12) fm) [18].} \tag{55b}$$

According to eq. (53), we have to add 69.1 kHz to the values (55a-b). Hence we obtain

$$(L_b)_{1S} = 8\,172\,732(14) \text{ kHz } (r_p\text{=0.805(11) fm), and} \tag{56a}$$

$$(L_b)_{1S} = 8\ 172\ 880(14) \text{ kHz (for } r_p=0.862(12) \text{ fm)}. \tag{56b}$$

Using the parameterization (43) with the corrected data (56a-b), we find the coefficients $(A_b)_{1S}$ and $(B_b)_{1S}$ as follows: $(A_b)_{1S} = 8\ 171\ 723.3$ kHz, $(B_b)_{1S} = 1\ 557.58$ kHz/fm. This allows us to determine the proton charge radius, equating to each other the corrected theoretical value and the experimental value (54). This coincidence occurs at

$$r_p = 0.846(22) \text{ fm}. \tag{57}$$

This estimation again well agrees with the value of $r_p$ determined through the 2S-2P Lamb shift in PBFT (48) and with the proton size determined in ref. [2].

**4. - Decay rate of bound muon**

In this section we return to the one-body problem for simplicity and recall that solving the Dirac-Coulomb equation (38) of [1], we applied the replacement

$$r = r'/b_n \gamma_n \tag{58}$$

(eq. (39) of [1]), which allowed us to present the Hamiltonian $\hat{H}$ in the form $\hat{H} = \hat{H}_{Schr} + \hat{V}$, where $\hat{H}_{Schr}$ is the non-relativistic Schrödinger Hamiltonian, whereas $\hat{V}$ is the perturbation.

Now we advance a hypothesis as follows: the transformation (58) represents not only a convenient mathematical trick, but eventually has a certain physical meaning. First we elaborate on the essence of our hypothesis in the classical language. Namely, we assume that the transformation (58) describes the change of radial scale for the particle, rotating in a vicinity of a host charge, as the function of their static binding energy (entering via the factor $b_n$), and orbital velocity (via the factor $\gamma_n$). This effect looks similar to the metric change in the gravitation field, where, the spatial scale depends on the gravitation potential (in the approximation of weak fields). Thus our hypothesis implies that the electric field also may change the metrics of space-time, though we assume that this effect is specific for bound wave-like particles and, in general, is not extended to the classical phenomena.

Further on we involve the undoubted statement that the light velocity in vacuum $c$, as measured by the distant observer, is not altered by the EM field. In addition, along the light pulse the space-time interval $S=0$. Hence, as viewed by the distant observer, $dS = c^2 dt^2 - dr^2 = 0$, and the same holds true for the bound particle: $dS' = c'^2 dt'^2 - dr'^2$. Since $c' = c$, then

$$dt' = dr'/c = dt/b\gamma. \tag{59}$$

Eq. (59) determines a unit time interval for the particle, and it shows that $dt'$ is larger than $dt$, since in the bound state $b\gamma < 1$. This means that the time rate is slow down for the particle in a bound state, as assessed by the distant observer.

In the quantum domain, eq. (59) acquires the form

$$dt' = \frac{dt}{b_n \gamma_n} = \frac{dt}{\left(1 - (Z\alpha/n)^2\right)^{1/2}}. \tag{60}$$

Herein $dt$ is interpreted as the time interval for a laboratory (macroscopic) observer.

Eq. (60) implies that the time rate for the bound micro-particle is not a constant value, but it varies with the change of $n$ and $Z$.

One should emphasize that the assumed effect (60) is independent of the conventional relativistic dilation of time for moving particles, and thus it can be subjected to experimental test separately. As an example, below we consider the decay rate for the bound muon in the atoms with $n=1$ (1S state) and various Z, and compare the results of calculations with the available experimental data.

It is known that a dominant channel of a decay of free muon is

$$\mu^- \to e^- + \tilde{\nu}_e + \nu_\mu \tag{61}$$

with the rate $\tau_0 = 2.2 \cdot 10^{-6}$ s in its proper reference frame.

Negative muons being captured by atoms, must necessarily be in a bound 1S state. Such a muon disappears by two competing processes: nuclear capture and decay (61). The cross-section of nuclear capture rapidly increases with $Z$; however, using the "start-stop" technique with registration of electrons in the reaction (61), one can measure the rate of this reaction separately.

To the moment, there were known three effects which make the rate of (61) different for bound muon than for free muon: phase space effect, relativistic dilation of time and the electron Coulomb effect [21]. For light atoms, the first and third effects almost eliminate each other, and the relativistic dilation of time prevails. For heavy atoms the situation becomes more complicated. The author of the mentioned paper [21] has made numerical calculations of bound muon decay rate $\tau_b$ versus $Z$ and plotted the corresponding curve to be shown in Fig. 1 (upper curve). In the same figure we show the experimental results [22], which drastically deviate from the theoretical curve especially at large $Z$.

In order to explain this deviation, Huff paid the attention to the substantial difference between the electron spectra for bound muon decay and that for free muon decay [21]. Hence the decay rate ratio $\tau_b/\tau_0$, before comparing with the experimental results, must be corrected for two effects: 1) the energy threshold for detection of the decay electrons; 2) the energy loss by the decay electrons in the target. By this way Huff has corrected the computed value of $\tau_b/\tau_0$ for iron ($Z=26$), antimony ($Z=51$), tantalum ($Z=73$) and lead ($Z=82$), which are depicted in Fig. 1 by hollow circles. Nonetheless, the deviation between experimental data and corrected theoretical values remains appreciable.

Now we assume that the remaining deviation of theoretical and experimental data reflects the effect of additional dilation of time for the bound muon according to eq. (60). Since this effect has a general character and depends on the binding energy of muon, it cannot be mixed with other effects influencing the decay rate of bound muon, mentioned above. Therefore, in order to take into account the time transformation (60) for the bound muon, we have to multiply Huff's data by factor $\left(1-(Z\alpha)^2/n^2\right)^{1/2}$. The results of our corrections are shown in Fig. 1 as triangles. One can observe a very satisfactory agreement of our results with the experimental data of [22].

It would be fair to add that the second author has developed and applied long ago, his ideas (which, in effect, initiated the present work), to predict the bound muon decay rate retardation. He postulated that the rest mass of any object bound to a given field, owing to the law of relativistic energy conservation, must be decreased as much as the mass equivalent of the static binding energy coming into play (and this already at rest, when macroscopic objects are considered) [23-26]. Along this way he also arrived to the variation of time rate for the bound particle as a function of the binding energy and obtained for bound muon a theoretical curve similar to presented by us in Fig. 1.

## 5. - Conclusion

In this paper we verify the Pure Bound Field Theory (PBFT) based on re-postulated Dirac and Breit equations, at the scale of hyperfine contributions to the atomic energy levels. On this way we consistently considered the fine structure corrections, corrections to hyperfine spin-spin interaction for the leptonic atoms, as well as the corrections to the Lamb shift.

The corrections of PBFT to common results of atomic physics stem from the replacements (5a-e), which reflect the modification of the structure of bound EM field for the system of bound charged with the forbidden radiation within the total momentum conservation constraint, when the wave equation (12) of ref. [1] for the operator of vector potential is replaced by the Poisson-like equation (14) of [1].

We have demonstrated that the fine structure corrections have the order of magnitude $mc^2(Z\alpha)^6 m/M$ and scales as $n^{-5}$ and $n^{-6}$. Hence they are practically significant only for 1S state of hydrogenlike atoms, in particular, in the re-estimation of ground state Lamb shift in the hydrogen atom. For 1S-2S interval in positronium, there appears an additional component of cor-rection due to the appropriate modification of annihilation term in PBFT, and both corrections completely eliminate the available discrepancy between theoretical value and experimental data for 1*S*-2*S* interval.

The corrections brought by PBFT to the spin-spin interval occur negligible for the hydrogenlike atoms with $m \ll M$ (for hydrogen and heavier atoms, it is due to the negligible value of the correcting factor $(Z\alpha)^2 \frac{2mM}{(M+m)^2} W_{s-s}$ in comparison with calculation uncertainty [1]; for muonium, eq. (29), it is due to cancellation by the same PBFT correction in the magnetic moment to mass ratio value derived from Zeeman effect). At the same time, such PBFT correction acquires significant value for the spin-spin interval of 1*S* state of positronium (eq. (30)), where the correction of PBFT removes the discrepancy between theoretical and experimental results.

The PBFT correction to the Lamb shift emerges due to the replacements (5a-e) in corresponding QED equations. Such a correction is directly applicable to the classic 2*S*-2*P* Lamb shift (eqs. (33)), where for the hydrogen atom we get an exact coincidence of corrected by us theoretical value of proton charge radius $r_p$=0.841(6) fm (eq. (48)) with the latest result of measurement via the 2*S*-2*P* Lamb shift in muonic hydrogen $r_p$=0.84184(67) fm [2]. The obtained correction to 2*S*-2*P* Lamb shift contributes to the corresponding correction to 1*S* Lamb shift via eq. (37), along with the correction of PBFT to 1*S*-2*S* interval (20b). Introducing the PBFT corrections to the 1*S* Lamb shift in hydrogen, we obtained the proton charge radius $r_p$=0.846(22) fm (eq. (57)), which practically coincides with the result yield by the classic 2*S*-2*P* Lamb shift.

In Table 1 we summarize the results of QED without and with the corrections we introduced, in comparison with corresponding experimental data. These data completely support our principal idea to modify the Dirac equation for non-radiative EM field of bound electron, which, in its turn, lead to further modifications of equations of the atomic physics. These modifications induce corrections into the effects to be not directly related to each other, but characterized by the same final result: practical elimination of deviations between theory and experiment.

We emphasize that the fine structure correction of PBFT in positronium and corrections to spin-spin interaction result from the appropriate modification of Breit equation in PBFT. In the analysis of radiative corrections to the atomic energy levels, we consider PBFT as a complementary to QED, and the modifications of QED core structure are not implied, excepting the replacements (5a-e) in the input of QED expressions.

Further, we have predicted a novel phenomenon for the bound micro-particles: a variation of their time rate as the function of their binding energy, which occurs independently on the standard relativistic dilation of time for moving objects. Perhaps, the assumed variation of time rate with a binding energy explains the result of our recent experiment on the Mössbauer effect in a rotating system [27], which also shows a frequency/energy shift additional to relativistic time dilation.

Involving this effect into the analysis of bound muon decay rate $\tau_b$ in muonic atoms, we have reached the quantitative agreement between theoretical and experimental data describing the dependence of $\tau_b$ on the atomic number *Z*. At the same time, one should mention that the experimental data on $\tau_b(Z)$ obtained in [22] and [28] contradict each other in some points, and new precise experiments for direct measurement of $\tau_b(Z)$ dependence would be of a high importance.

Further on we emphasize the universal character of PBFT, which is also applied to the analysis of heavy atoms, where we also achieve the promising results. Such an analysis will be done in a separate contribution.

Finally, we believe that the present contribution will stimulate further detailed experimental research in the atomic physics.

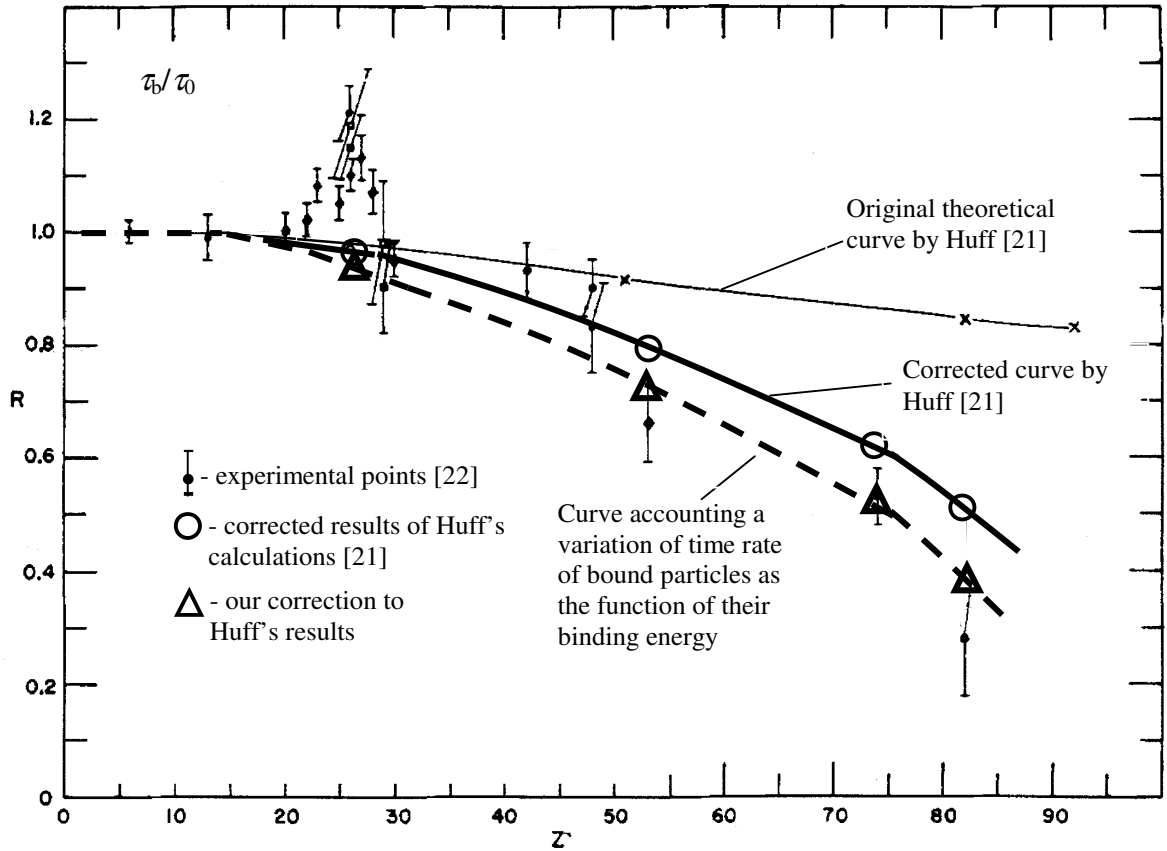

Fig. 1. Comparison of the results of theoretical calculation of the bound muon decay rate versus $Z$ with the experimental data of [22]. The peak near $Z=26$ observed in [22] was not confirmed in later experiment [27].

Table 1

Correction of QED results within PBFT in comparison with corresponding experimental (recommended) values for those parameters, where a high measuring precision has been achieved

| Parameter | QED result | Result corrected by us | Experimental value(s) | Ratio "deviation/uncertainty" before correction | Ratio "deviation/uncertainty" after correction |
|---|---|---|---|---|---|
| 1S-2S interval in positronium | 1 233 607 222.2(6) MHz | 1 233 607 214.7(6) MHz (sub-section 3.1, eq. (40)) | 1 233 607 216(1) MHz [12] | 3.0· | 1 |
| Spin-spin splitting in positronium | 203 391.7(8) MHz | 203 386(1) (sub-section 3.2, eq. (42)) | 203 389(2) [13] 203 387(2) [14] | 1.5 2.5 | -1.5 <1 |
| Spin-spin splitting in muonium | 4 463 302.88(55) | remains non-corrected (sub-section 2.2, eq. (29)) | 4 463 302.78(5) [5] | <1 | <1 |
| Proton charge radius (2S-2P Lamb shift in hydrogen) | 0.876(6) fm | 0.841(6) fm (sub-section 3.3, eq. (48)) | 0.84184(67) fm [2] | 5.7 | <1 |
| Proton charge radius (1S Lamb shift in hydrogen) | 0.876(6) fm | 0.846(22) fm (sub-section 3.5, eq. (57)) | 0.84184(67) fm [2] | 5.7 | <1 |